# Growth and Investigation of Nd$_{1-x}$Sm$_x$ScO$_3$ and Sm$_{1-x}$Gd$_x$ScO$_3$ Solid-Solution Single Crystals

R. Uecker[a], D. Klimm[a], R. Bertram[a], M. Bernhagen[a], I. Schulze-Jonack[a], M. Brützam[a], A. Kwasniewski[a], Th. M. Gesing[b] and D.G. Schlom[c,d]

[a]Leibniz Institute for Crystal Growth, Max-Born-Str. 2, D-12489 Berlin, Germany

[b]Solid State Chemical Crystallography, Institute for Inorganic Chemistry, University Bremen, Leobener Straße /NW2, D-28359 Bremen, Germany

[c]Department of Materials Science and Engineering, 230 Bard Hall, Cornell University, Ithaca, New York 14853-1501, USA

[d]Kavli Institute at Cornell for Nanoscale Science, Ithaca, New York, 14853, USA

Abstract

The pseudo-cubic lattice parameters of rare-earth (*RE*) scandate, *RE*ScO$_3$, single crystals grown by the Czochralski technique with *RE*=Dy to Pr lie between about 3.95 and 4.02 Å. These crystals are the only available perovskite substrates in this lattice constant range that can withstand virtually any thin film growth conditions. Two members of this series, PmScO$_3$ and EuScO$_3$, are, however, not suitable for substrate applications. Because the pseudo-cubic lattice parameters between neighbouring *RE*ScO$_3$ compounds decrease with rising atomic number of the *RE* in about 0.01 Å steps, the unsuitability of PmScO$_3$ (radioactivity) and EuScO$_3$ (incompatibility with Si) causes an interruption in this lattice spacing sequence. To replace them, solid solutions of their adjacent rare-earth scandates, i.e., (Nd$_{0.5}$Sm$_{0.5}$)ScO$_3$ and (Sm$_{0.5}$Gd$_{0.5}$)ScO$_3$, were grown by the Czochralski method. Their average pseudo-cubic lattice parameters of 3.9979 Å and 3.9784 Å are very close to those of PmScO$_3$ and EuScO$_3$, respectively, and they show very low segregation. These qualities make these solid solutions excellent substitutes for PmScO$_3$ and EuScO$_3$.

1. Introduction

Over the past decade perovskite type rare-earth (*RE*) scandate single crystals *RE*ScO$_3$ with *RE* = Dy - Pr have become widely used substrates for the growth of perovskite thin films[1,2,3]. These *RE*ScO$_3$ substrates have pseudo-cubic lattice parameters covering the range between 3.95 and 4.02 Å. The only other perovskite substrate with a lattice parameter in this range is KTaO$_3$ (3.989 Å). An issue with KTaO$_3$, however, is that it can only be grown in small dimensions from self-fluxes[4] and tends to dissociate under typical growth conditions used for the growth of perovskite thin films (substrate temperatures ~750 °C) due to potassium evaporation[5]. Following Goldschmidt's tolerance factor rule[6] only *RE* scandates between HoScO$_3$ and LaScO$_3$ exhibit the perovskite type structure[7], i.e., they are isomorphous. The orthorhombic structure of these *RE* scandates (standard setting space group *Pnma*, no. 62) is a distorted derivative of the cubic perovskite structure, where the *RE* occupy the eightfold coordinated *A* sites and Sc the centres of the [ScO$_6$] octahedra (Figure 6, inset). The unit cell contains four formula units[7].

The crystals ranging from DyScO$_3$ to PrScO$_3$ are grown by the conventional Czochralski technique[8,9]. The most recent member to be grown as a single crystal, PrScO$_3$[10], is the one with the highest melting temperature. Of the missing *RE* scandates, CeScO$_3$ and LaScO$_3$, (the largest *RE* scandates) cannot be grown by the conventional Czochralski method due to their exceptionally high melting temperatures[11] exceeding the maximum load of an iridium crucible, which is about 2250° C. The smallest *RE* scandate, HoScO$_3$, cannot be prepared because it is not a stable phase at its melting point. In the vicinity of its melting temperature the stable polymorph of HoScO$_3$ has the fluorite structure. Only about 200 °C below its melting point does the perovskite polymorph of HoScO$_3$ become stable[11,12]. Additionally, two further members of the above range of *RE* scandates are problematic as substrates for the growth of thin films: (1) PmScO$_3$ because it is radioactive and (2) EuScO$_3$, which due to the multivalent nature of Eu is thermodynamically unstable in direct contact with silicon[13]. Because the pseudo-cubic lattice parameters of the *RE* scandates neighbouring PmScO$_3$ and EuScO$_3$ differ from them by about 0.01 Ångstrom, the lack of single crystal substrates of these compounds results in an interruption of the lattice parameters continuity, i.e., the set of *RE* scandates substrate crystals which could enable strain engineering of perovskite thin films in regular 0.01 Å steps is incomplete (Figure 1).

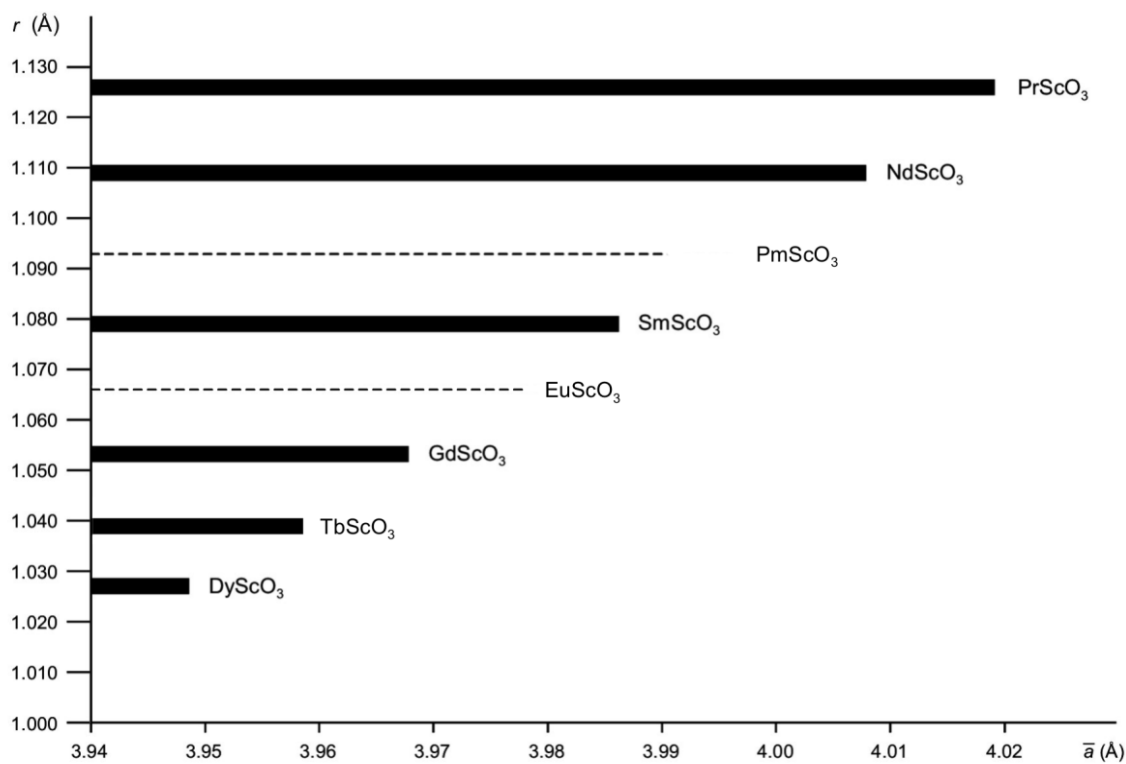

Figure 1: Relation between ionic radii of the *RE* ions in eightfold coordination ($r$)[14] and the pseudo-cubic lattice parameter ($\bar{a}$) of the corresponding *RE* scandates. The pseudo-cubic lattice constant of PmScO$_3$ is calculated from the data in Weigel *et al.*[15].

In the present report we investigated whether solid solutions (SSLs) utilizing *RE*s adjacent to these missing *RE* scandates can be grown, i.e., whether PmScO$_3$ could be replaced by a NdScO$_3$ - SmScO$_3$ SSL and EuScO$_3$ by a SmScO$_3$ - GdScO$_3$ SSL. Based on Vegard`s rule both solid solutions should have nearly the same lattice parameters as the material replaced if prepared in a 1:1 molar composition ratio. Provided segregation is minimal, such solid solutions could be valuable substitutes for PmScO$_3$ and EuScO$_3$ single crystal substrates, enabling strain engineering of thin films grown upon them.

Because all *RE* nearest neighbours in the range $DyScO_3$-$PrScO_3$ have isomorphous crystal structures and their corresponding atoms have similar ionic radii, the only missing precondition for the possibility to grow solid solutions is the question of congruent melting of the end members. $DyScO_3$, $GdScO_3$, $SmScO_3$, $NdScO_3$, and $LaScO_3$ were reported to melt congruently[11]; our DTA measurements on $TbScO_3$ and $PrScO_3$ show that both compounds also melt congruently at 2095° C and 2197° C, respectively.

If the formation of solid solutions between all nearest neighbours in the range Dy - Pr is possible, a route to continuous strain engineering of perovskite thin films and fine-tuning of film properties could be opened.

2. Experimental

2.1. Crystal growth

The starting oxides were of a minimum of 99.99% purity. To prepare the starting melt the powders were dried, mixed according to the formula $Sm_{0.5}Gd_{0.5}ScO_3$ and $Nd_{0.5}Sm_{0.5}ScO_3$, sintered at 900-1100° C, and finally isostatically pressed.

The crystals were grown by the conventional Czochralski technique with RF-induction heating (25 kW generator) and automatic diameter control. Because of the high melting temperatures of the *RE* scandates, iridium crucibles with an active afterheater must be used. The crucible dimensions were 40 mm in diameter and height; the afterheater was of the same diameter and 80 mm in height. To suppress the frequently observed spiral formation of several *RE* scandates, a conical baffle was placed directly on the crucible[8]. Thermal insulation was provided by an outer alumina ceramic tube filled with zirconia felt and granules. The crystals were grown under flowing nitrogen or argon. To grow $NdScO_3$-$SmScO_3$ and $SmScO_3$-$GdScO_3$ SSLs, a seed crystal from the *RE* scandate end member with the higher melting point was used. In the case of $Sm_{0.5}Gd_{0.5}ScO_3$ the pulling rate was 1.0 mm $h^{-1}$ and the rotation rate was 10 rpm. $Nd_{0.5}Sm_{0.5}ScO_3$ was grown with a pulling rate of 0.5 mm $h^{-1}$ and a rotation rate of 15 rpm. Both solid solutions were grown along the [101] direction to enable efficient preparation of (101) substrates. The rectangular surface mesh of the (101) plane (which corresponds to a pseudo-cubic (100) surface) is preferred for the epitaxy of (100)-oriented pseudo-cubic perovskite thin films.

2.2. Crystal characterization

2.2.1. Crystal composition

The chemical composition of the crystals was investigated with an inductively coupled plasma - optical emission spectrometer (ICP-OES) IRIS Intrepid HR Duo. The samples were prepared by microwave digestion with $HNO_3$ (240° C, 20 min).

2.2.2. Melting behaviour, melting point and heat of fusion

Differential thermal analysis (DTA) was performed with a NETZSCH STA429, where a tungsten mesh heater in a 99.9999% pure He atmosphere allows measurements up to 2400° C with a heating rate of 15 K/min. Typical samples consisted of 20 – 50 mg of crushed pieces from the grown $RE$ScO$_3$ single crystals. Prior measurements indicate that the congruently melting composition of all $RE$ scandates investigated to date occur at $RE$:Sc ratios slightly smaller than unity[9]. In addition to pure $GdScO_3$, $SmScO_3$, and $NdSmO_3$ compounds, several mixtures of $GdScO_3$ - $SmScO_3$ and $NdScO_3$ - $SmScO_3$ were also measured. The samples were placed in tungsten crucibles with lids.

The temperatures in the DTA were measured with W/Re thermocouples. The indicated temperature was observed to drift at the very high temperatures ($T > 2000°$ C) used in the present study. This drift was accounted for by occasional measurements of $Al_2O_3$ as a reference material with known melting point and heat of fusion ($T_f = 2054°$ C, $\Delta H_f = 118.4$ kJ/mol)[16]. The temperature error in the range beyond 2000° C was estimated to be ±15 K.

2.2.3. Lattice parameters

The lattice parameters of the solid solutions were calculated from X-ray powder diffraction data measured with a Seiffert UDR6 diffactometer (CuK$_\alpha$ radiation; 2θ range 20° - 70°).

3. Results

3.1. Grown solid solutions

Two examples of the solid solutions grown by the Czochralski method are shown in Figure 2.

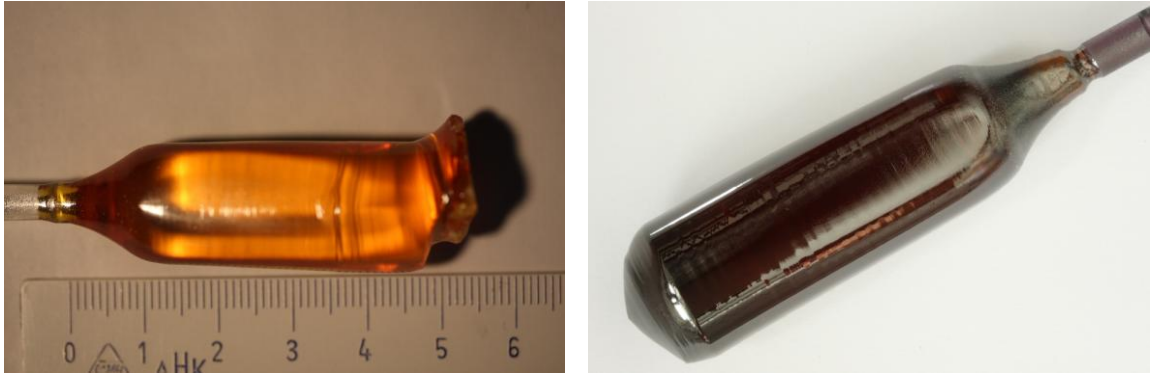

Figure 2: $Sm_{0.5}Gd_{0.5}ScO_3$ (left) and $Nd_{0.5}Sm_{0.5}ScO_3$ (right) solid-solution single crystals

The cylindrical $Sm_{0.5}Gd_{0.5}ScO_3$ crystal was about 17 mm in diameter and 40 mm in length (Figure 2 left). At the tail end of the crystal the beginning of spiral formation can be seen. The colour of the crystal is determined by the colour of the trivalent *RE* ions, i.e., the colourless $GdScO_3$ and the yellow $SmScO_3$ lead to the yellow colour of this solid solution.

The diameter of the $Nd_{0.5}Sm_{0.5}ScO_3$ crystal was about 18 mm and its cylindrical length was about 50 mm (Figure 2 right). Here the violet colour of $NdScO_3$ and the yellow colour of $SmScO_3$ led to a brownish colour of the solid solution.

Both solid solutions form four facets at the periphery: the $(12\bar{1})$, $(\bar{1}21)$, $(1\bar{2}\bar{1})$, and $(\bar{1}\bar{2}1)$ planes. The corresponding facets have not been observed at the crystal-melt interface.

X-ray rocking curves taken in a triple-axis geometry on a Rigaku SmartLab (utilizing the 220 reflections of Ge monochromator crystals on both the receiving and diffracted side) revealed both solid solutions to have good perfection. The rocking curve full width at half maximum (FWHM) for both solid-solution single crystals was 12 arc sec in $\omega$ (Figure 3).

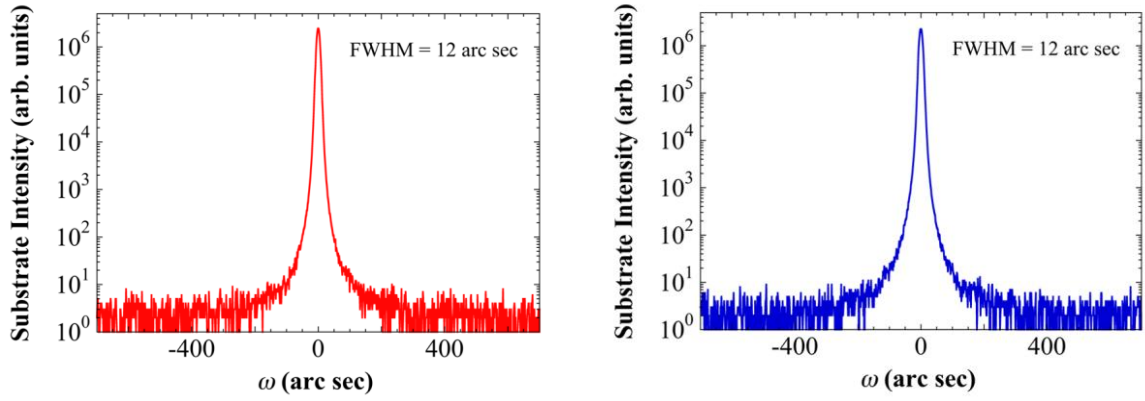

Figure 3: Rocking curves of the $Gd_{0.5}Sm_{0.5}ScO_3$ 220 peak (left) and the $Nd_{0.5}Sm_{0.5}ScO_3$ 220 peak (right). Both have FWHM of 12 arc sec (0.0033°).

The dislocation density is about $2 \times 10^4$ cm$^2$. They show neither stress birefringence nor scattering centers.

3.2. Crystal characterization

3.2.1. Chemical composition

The chemical composition of the crystals measured by inductively coupled plasma-optical emission spectrometry is shown in Tables 1a and 1b.

| $Gd_{0.5}Sm_{0.5}ScO_3$ | $Gd_2O_3$ (ma %) | $Sm_2O_3$ (ma %) | $Sc_2O_3$ (ma %) | $Gd_2O_3$ (mol %) | $Sm_2O_3$ (mol %) | $Sc_2O_3$ (mol %) |
|---|---|---|---|---|---|---|
| initial melt composition | 36.73 | 35.33 | 27.94 | 25 | 25 | 50 |
| crystal top | 35.53 ± 0.04 | 35.36 ± 0.04 | 29.02 ± 0.09 | 23.92 ± 0.03 | 24.74 ± 0.03 | 51.34 ± 0.16 |
| crystal bottom | 36.13 ± 0.24 | 35.47 ± 0.18 | 29.17 ± 0.12 | 24.14 ± 0.16 | 24.64 ± 0.13 | 51.23 ± 0.21 |

Table 1a: Chemical composition of the solid solution grown from a $Gd_{0.5}Sm_{0.5}ScO_3$ melt.

| $Nd_{0.5}Sm_{0.5}ScO_3$ | $Nd_2O_3$ (ma %) | $Sm_2O_3$ (ma %) | $Sc_2O_3$ (ma %) | $Nd_2O_3$ (mol %) | $Sm_2O_3$ (mol %) | $Sc_2O_3$ (mol %) |
|---|---|---|---|---|---|---|
| initial melt composition | 35.01 | 36.29 | 28.70 | 25 | 25 | 50 |
| crystal top | 36.69 ± 0.09 | 33.08 ± 0.07 | 29.56 ± 0.04 | 26.07 ± 0.06 | 22.68 ± 0.05 | 51.25 ± 0.07 |
| crystal bottom | 35.74 ± 0.02 | 35.03 ± 0.02 | 29.83 ± 0.04 | 25.11 ± 0.01 | 23.75 ± 0.01 | 51.14 ± 0.07 |

Table 1b: Chemical composition of the solid solution grown from a $Nd_{0.5}Sm_{0.5}ScO_3$ melt.

On the basis of the site occupancy refinements of the *RE* scandate single crystals which show a *RE*-deficiency on the *A*-site and O-vacancies on the oxygen position[17] the final crystallochemical formula of both solid solutions can be written as

top: $^A(\square_{0.052}Gd_{0.466}Sm_{0.482})^B ScO_{2.922}$

bottom: $^A(\square_{0.048}Gd_{0.471}Sm_{0.481})^B ScO_{2.928}$

and

top: $^A(\square_{0.049}Nd_{0.509}Sm_{0.442})^B ScO_{2.927}$

bottom: $^A(\square_{0.045}Nd_{0.491}Sm_{0.464})^B ScO_{2.933}$.

### 3.2.2. Thermal analysis and phase diagrams

Although reliable $\Delta H_f$ measurements are not possible when using a DTA sample carrier, we estimated $\Delta H_f$ with an accuracy of about 20% by comparing the melting peak areas of the *RE* scandates with that of $Al_2O_3$. An almost ideal behaviour of both the liquid (melt) phase and of the solid solutions are expected because the corresponding quasi-binary partners Gd/Sm and Nd/Sm are always second-nearest neighbours in the *RE* series and the phase equilibria were investigated at very high *T* (exceeding 2000° C). In this case, the solidus and liquidus of the systems can be calculated with the Schröder-van-Laar equation (1)

$$x^{sol} = \frac{\exp[a]-1}{[\exp[a]-\exp[b]]}$$
$$x^{liq} = \exp[b] \cdot x^{sol} \quad (1)$$
$$\text{where} \quad a,b = \frac{\Delta H_f}{R}\left(\frac{1}{T}-\frac{1}{T_f}\right) \text{ for both components}$$

from the experimental data measured for both components (Table 2)[18].

|  | $T_f$ (K) | $\Delta H_f$ (kJ/mol) |
|---|---|---|
| $GdScO_3$ | 2423 | 85 |
| $SmScO_3$ | 2440 | 105 |
| NdScO$_3$ | 2491 | 89.2 |

Table 2: Melting points (error ±15 K) and heat of fusion (error estimated to 20%) as determined by DTA measurements for the pure $RE$ScO$_3$ end members of this study.

The resulting liquidus and solidus are shown in the phase diagrams in Figure 4.

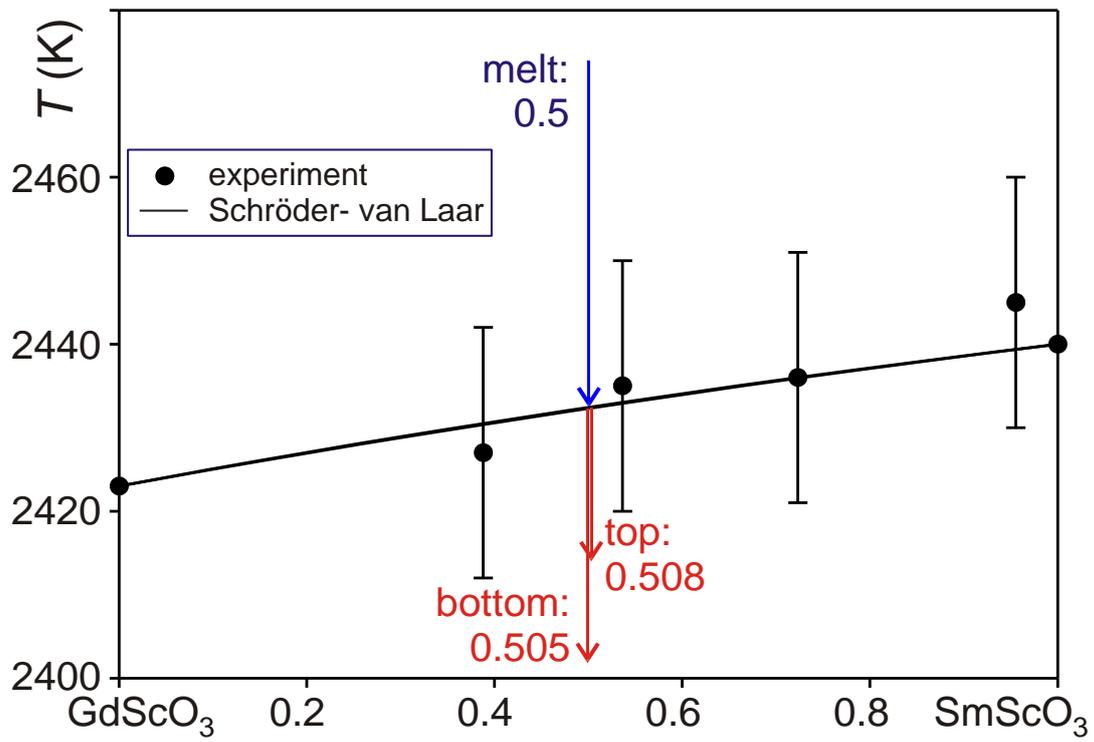

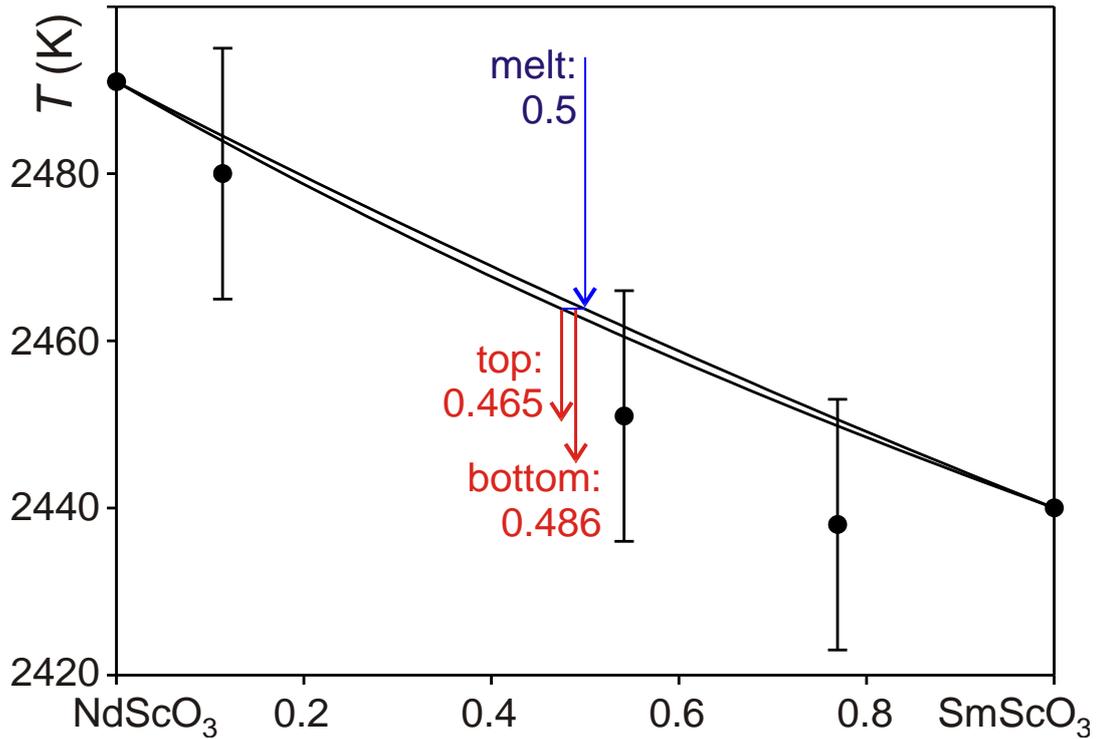

Figure 4: Experimental phase diagrams GdScO$_3$-SmScO$_3$ (top) and NdSmO$_3$-SmScO$_3$ (bottom) as derived from DTA melting peak onsets compared with theoretical liquidus and solidus curves calculated from $T_F$ and $\Delta H_f$. The arrow labels refer to the crystal growth data in Table 3.

| melt comp. | crystal | |
|---|---|---|
| | top | bottom |
| Gd$_{0.5}$Sm$_{0.5}$ScO$_3$ | Gd$_{0.492}$Sm$_{0.508}$ScO$_3$ | Gd$_{0.495}$Sm$_{0.505}$ScO$_3$ |
| Nd$_{0.5}$Sm$_{0.5}$ScO$_3$ | Nd$_{0.535}$Sm$_{0.465}$ScO$_3$ | Nd$_{0.514}$Sm$_{0.486}$ScO$_3$ |

Table 3: Segregation during the growth of Gd$_{0.5}$Sm$_{0.5}$ScO$_3$ and Nd$_{0.5}$Sm$_{0.5}$ScO$_3$ SSL`s.

Figure 4 compares the liquidus and solidus lines that are given by eq. (1) with experimental points (DTA melting peak onsets) that were obtained for several compositions within both systems.

It is obvious that the simplified model of ideal solutions for liquid and solid phases (Gd$_{1-x}$Sm$_x$)ScO$_3$ and (Nd$_{1-x}$Sm$_x$)ScO$_3$ describes the DTA results reasonably. It turns out that the 2-phase region (liquid + solid) is narrow in both cases – especially for the Gd-Sm system where the melting points differ less than in the Nd-Sm system (17 K or 51 K, respectively). Also the comparably small $\Delta H_f$ values given in Table 2 contribute to the tightness, which is responsible for the low segregation that is observed during Czochralski crystal growth of both systems at intermediate compositions $x$ = 0.5 (Table 3). As expected, the minor segregation leads always to a pronounced crystallization of the higher melting component at the top of the growing crystal, leaving a slightly depleted melt. The segregation reaches a maximum value 3.5 % in the Nd-Sm system, as the 2-phase region is there somewhat broader when compared with the Gd-Sm system.

The specific heat capacity $c_p(T)$ of pure single crystals (60 - 110 mg) was measured twice with dynamic DSC in a NETZSCH STA 449C "Jupiter", using a DSC-$c_p$ sample holder and platinum crucibles with lids. The measurements were performed according to ASTM E1269, in oxygen flow with a sapphire disc as a standard and a heating rate of 20 K/min. The results are shown in Figure 5, and fit parameters for the $c_p(T)$ functions are given in Table 4. It should be noted that heat flow calorimetry does not reach the high accuracy that is obtained, e.g., with drop-in calorimeters. The data shown in Figure 5 are the smoothed average of two subsequent measurements, and these subsequent runs differed typically by 2 %. The total error in these measurements, however, is estimated to be around 10 %.

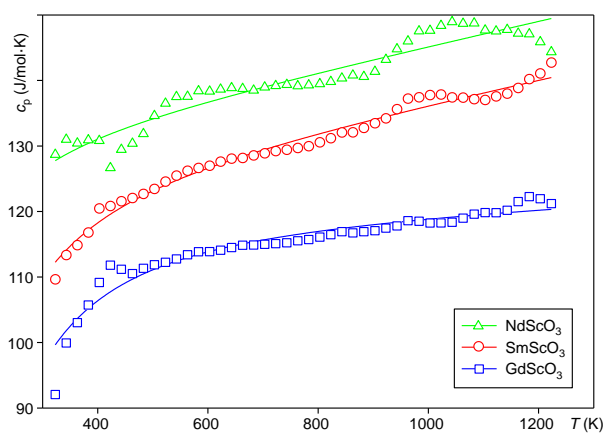

Figure 5: Specific heat capacity functions $c_p(T)$ obtained by dynamic DSC (symbols), together with fitted functions $c_p = a + bT + c/T^2$ ($T$ in Kelvin). For parameters see Table 4.

| crystal | $a$ | $b$ | $c$ |
|---|---|---|---|
| NdScO$_3$ | 127.00 | 0.0187 | $-5.497409\times10^{-5}$ |
| SmScO$_3$ | 119.99 | 0.0175 | $-1.397454\times10^{-6}$ |
| GdScO$_3$ | 116.86 | $3.8854\times10^{-3}$ | $-1.927588\times10^{-6}$ |

Table 4: Parameters of the specific heat capacity functions $c_p(T) = a + bT + c/T^2$ of the three rare earth scandates used in this study with $T$ in Kelvin and $c_p$ in J/(mol·K).

3.2.3. Lattice parameters

Due to the low segregation, the composition of crystal top and bottom of both SSLs differs by less than 2 mol-%, i.e., the lattice parameters vary only very little over the crystal length. While investigating the SSLs, the lattice parameters of all $RE$ scandate single crystals were calculated from X-ray powder diffraction data and are given in Table 5. These data are of basic interest for epitaxy.

| $RE$ | orthorhombic lattice parameters (Å) | | | | | | pseudo-cubic lattice parameters (Å) | | |
|---|---|---|---|---|---|---|---|---|---|
| | $a$ | $b$ | $c$ | $da$ | $db$ | $dc$ | $0.5\sqrt{a^2+b^2}$ | $b/2$ | $\overline{a_{pc}}$ |
| Pr | 5.7797863 | 8.0244991 | 5.6066875 | 0.0000533 | 0.0000798 | 0.0000529 | 4.0262 | 4.0123 | 4.0193 |
| Nd | 5.7761026 | 8.0033835 | 5.5761705 | 0.0000495 | 0.0000721 | 0.0000491 | 4.0143 | 4.0017 | 4.0080 |
| Sm,Nd | 5.7685143 | 7.9843447 | 5.5537820 | 0.0000571 | 0.0000852 | 0.0000575 | 4.0038 | 3.9922 | 3.9980 |
| Sm | 5.7586120 | 7.9627969 | 5.5279832 | 0.0000530 | 0.0000795 | 0.0000532 | 3.9912 | 3.9814 | 3.9863 |
| Gd,Sm | 5.7543952 | 7.9490397 | 5.5065474 | 0.0000538 | 0.0000776 | 0.0000539 | 3.9823 | 3.9745 | 3.9784 |
| Gd | 5.7454058 | 7.9314013 | 5.4805002 | 0.0000551 | 0.0000834 | 0.0000551 | 3.9701 | 3.9657 | 3.9678 |
| Tb | 5.7299335 | 7.9164596 | 5.4638353 | 0.0000445 | 0.0000639 | 0.0000444 | 3.9587 | 3.9583 | 3.9585 |
| Dy | 5.7163901 | 7.9031349 | 5.4400236 | 0.0000414 | 0.0000628 | 0.0000406 | 3.9456 | 3.9516 | 3.9486 |

Table 5: Orthorhombic and pseudo-cubic lattice parameters of the *RE* scandates that have been grown in single crystal form, including the SSLs $Sm_{0.48}Nd_{0.52}ScO_3$ and $Gd_{0.49}Sm_{0.51}ScO_3$ ($RE = REScO_3$).

A comparison of the lattice parameters of both SSLs and the *RE* scandates that were targeted to replace ($EuScO_3$ and $PmScO_3$) is shown in Table 6.

| *RE* scandate | orthorhombic lattice parameters (Å) | | | pseudo-cubic lattice parameters (Å) | | |
|---|---|---|---|---|---|---|
| | $a$ | $b$ | $c$ | $0.5\sqrt{a^2+b^2}$ | $b/2$ | $\overline{a_{pc}}$ |
| $Gd_{0.49}Sm_{0.51}ScO_3$ | 5.7544 | 7.9490 | 5.5065 | 3.9823 | 3.9745 | 3.9784 |
| $EuScO_3$ | 5.7541 | 7.9474 | 5.5063 | 3.9821 | 3.9737 | 3.9779 |
| $Sm_{0.48}Nd_{0.52}ScO_3$ | 5.7685 | 7.9843 | 5.5538 | 4.0038 | 3.9922 | 3.9980 |
| $PmScO_3$[15] | 5.79 | 7.94 | 5.56 | 4.01 | 3.97 | 3.99 |

Table 6: Orthorhombic and pseudo-cubic lattice parameters of $EuScO_3$ and $PmScO_3$ compared with the substituting *RE* scandate solid solutions.

As can be seen from Table 6, the average pseudo-cubic lattice parameter $\overline{a_{pc}}$ of $Gd_{0.49}Sm_{0.51}ScO_3$ matches very well with the one of $EuScO_3$, while that of $Sm_{0.48}Nd_{0.52}ScO_3$ is not so close to the value determined from the published data for $PmScO_3$[15].

In Figure 6 the dependence of the orthorhombic lattice parameter of all grown *RE* scandates including both SSLs on the ionic radii of the *RE* ions ($RE^{3+}$ in eightfold coordination) is shown.

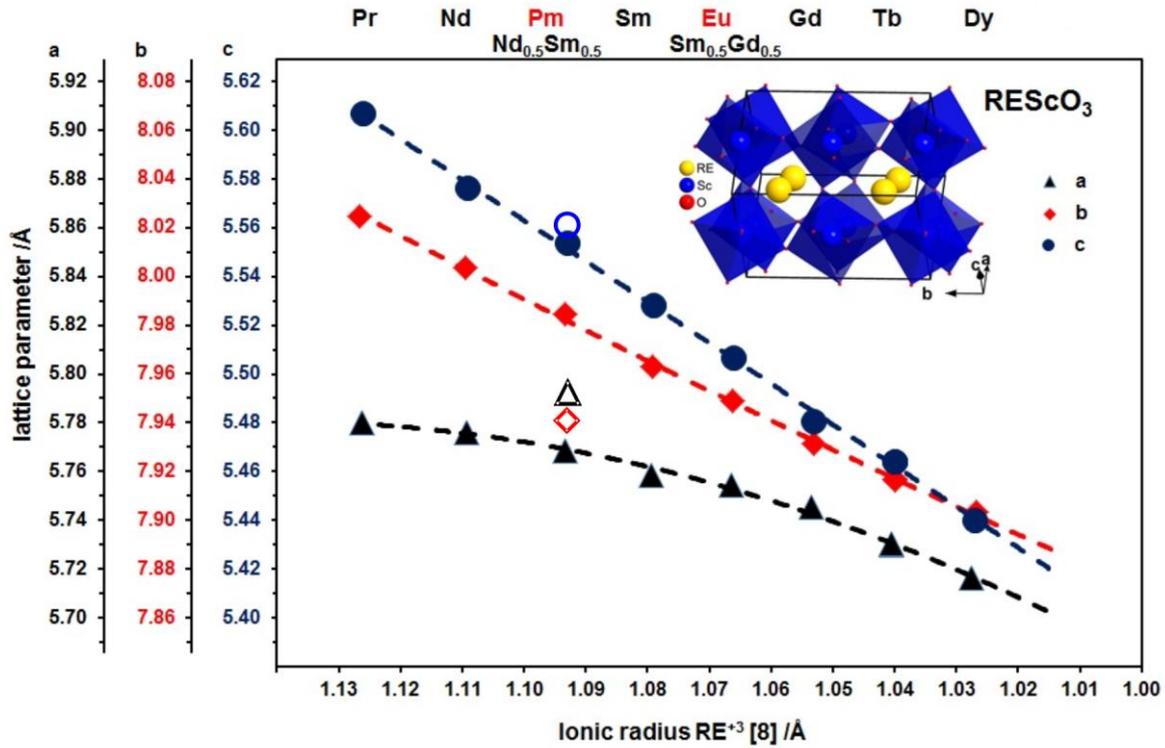

Figure 6: Dependence of the orthorhombic lattice parameters of the *RE* scandates on the ionic radii of the relevant *RE* ions in eightfold coordination ($RE^{3+}$)[14] (the empty symbols correspond to the lattice parameters of $PmScO_3$ determined from the published data[15]).

As seen in Figure 6 the orthorhombic lattice parameters of the *RE* scandates depend directly on the ionic radii of the *RE* elements. It also can be seen in Figure 6, however, that the lattice parameters of $Sm_{0.48}Nd_{0.52}ScO_3$ fit this dependence much better than those of $PmScO_3$. This is probably due to the low accuracy of the lattice constant estimation of $PmScO_3$ (PDF 033-1091, quality: B).

4. Conclusions

In order to replace $EuScO_3$ and $PmScO_3$, which have drawbacks for substrate applications for perovskite thin films compared to the surrounding *RE* scandates between $DyScO_3$ and $PrScO_3$, solid solutions of their adjacent rare-earth scandates were grown by the Czochralski method. To achieve this goal the phase diagrams of the systems $GdScO_3$-$SmScO_3$ and SmScO$_3$-NdScO$_3$ were determined by thermodynamic calculations and DTA measurements. Both systems show complete solubility and very low segregation. Additionally, the specific heat capacity of the pure single crystals was determined. The lattice parameters of Gd$_{0.49}$Sm$_{0.51}$ScO$_3$ and Sm$_{0.48}$Nd$_{0.52}$ScO$_3$ make them suitable substitutes for EuScO$_3$ and PmScO$_3$ substrates.

Based on these results it is expected that solid solutions between all neighbouring *RE* scandates can be formed, thus enabling substrates with continuously adjustable lattice parameters over the range of pseudo-cubic perovskite lattice spacings between 3.95 and 4.02 Å.

Acknowledgement:

The authors are very grateful to Che-Hui Lee at Cornell University for the triple-crystal X-ray diffraction measurements.

References:

[1] K.J. Choi, M. Biegalski, Y.L. Li, A. Sharan, J. Schubert, R. Uecker, P. Reiche, L.-Q. Chen, V. Gopalan, D.G. Schlom, C.B. Eom, Science **306**, 1005 (2004).
DOI:10.1126/science.1103218

[2] J.H. Haeni, P. Irvin, W. Chang, R. Uecker, P. Reiche, Y.L. Li, S. Choudhury, W. Tian, M.E. Hawley, B. Craigo, A.K. Tagantsev, X.Q. Pan, S.K. Streiffer, L.Q. Chen, S.W. Kirchoefer, J. Levy, D.G. Schlom, Nature **430**, 758 (2004).
DOI:10.1038/nature02773

[3] J.H. Lee, L. Fang, E. Vlahos, X. Ke, Y.W. Jung, L.F. Kourkoutis, J.-W. Kim, P.J. Ryan, T. Heeg, M. Roeckerath, V. Goian, M. Bernhagen, R. Uecker, P.C. Hamme, K.M. Rabe, S. Kamba, J. Schubert, J.W. Freeland, D.A. Muller, C.J. Fennie, P.E. Schiffer, V. Gopalan, E. Johnston-Halperin, D.G. Schlom, Nature **466**, 954 (2010).
DOI:10.1038/nature09331

[4] S. Zlotnik, P.M. Vilarinho, M.E.V. Costa, J.A. Moreira, A. Almeida, Crystal Growth & Design **10**, 3397 (2010) .
DOI:10.1021/cg100036v

[5] H.-M. Christen, L.A. Boatner, J.D. Budai, M.F. Chisholm, L.A. Gea, D.P. Norton, C. Gerber, M. Urbanik, Appl. Phys. Lett. **70**, 2147 (1997).
DOI:10.1063/1.119082

[6] V.M. Goldschmidt, "Die Gesetze der Krystallochemie," *Naturwissenschaften* **14**, 477 (1926).
DOI:10.1007/BF01507527

[7] R.P. Liferovich, R.H. Mitchell, J. Solid State Chem. **177**, 2188 (2004).

DOI:10.1016/j.jssc.2004.02.025

[8] R. Uecker, H. Wilke, D.G. Schlom, B. Velickov, P. Reiche, A. Polity, M. Bernhagen, M. Rossberg, J. Crystal Growth **295**, 84 (2006).

DOI:10.1016/j.jcrysgro.2006.07.018

[9] R. Uecker , B. Velickov, D. Klimm, R. Bertram, M. Bernhagen, M. Rabe, M. Albrecht, R. Fornari, D.G. Schlom, J. Crystal Growth **310**, 2549 (2008).

DOI:10.1016/j.jcrysgro.2008.01.019

[10] T.M. Gesing, R. Uecker, J.C. Buhl, Z. Kristallogr. NCS **224**, 365 (2009).

DOI:10.1524/ncrs.2009.0159

[11] J.M. Badie, Rev. Int. Hautes Temp. Refract., Fr. **15**, 183 (1978).

[12] J.M. Badie and M. Foex, J. Solid State Chem. **26**, 311 (1978).

DOI:10.1016/0022-4596(78)90165-2.

[13] K.J. Hubbard, D.G. Schlom, J. Mater. Res. **11**, 2757 (1996).

DOI:10.1557/JMR.1996.0350

[14] R.D. Shannon, Acta Crystallographica Section A **32**, 751 (1976).

DOI:10.1107/S0567739476001551

[15] F. Weigel, V. Scherer, Radiochim. Acta **7**, 50 (1967).

[16] FactSage 6.3, Thermodynamic software and database, www.factsage.com.

[17] B. Velickov, V. Kahlenberg, R. Bertram, M. Bernhagen, Z. Kristallogr. NCS **222**, 466 (2007).

DOI:10.1524/zkri.2007.222.9.466

[18] P. Paufler, Phasendiagramme, Akademie-Verlag, Berlin 1981 (in German).

DOI:10.1002/zamm.19820621217